\documentclass[12pt]{iopart}
\usepackage{hyperref}
\usepackage{graphicx}
\usepackage{iopams}
\usepackage{bbold}
\usepackage{xcolor}
\usepackage[xcolor]{changes}
\definechangesauthor[color=magenta]{NL}
\definechangesauthor[color=orange]{GS}
\definechangesauthor[color=blue]{VS}

\newcommand{\ket}[1]{|#1\rangle}
\newcommand{\bra}[1]{\langle#1|}

\setremarkmarkup{\textit{Remark #1:\textcolor{Changes@Color#1}{#2}}}

\begin{document}
\title[]{Entanglement protection of high-dimensional states by adaptive optics}

\author{Giacomo Sorelli$^1$, Nina Leonhard$^2$, Vyacheslav N. Shatokhin$^1$, 
Claudia Reinlein$^2$ and Andreas Buchleitner$^1$}
\address{$1$ Physikalisches Institut, Albert-Ludwigs-Universit\"at Freiburg, 
Hermann-Herder-Stra\ss{}e 3, D-79104 Freiburg, Germany}
\address{$2$ Fraunhofer Institute for Applied Optics and Precision 
Engineering, Albert-Einstein-Stra\ss{}e 7, 07745 Jena, Germany}


\begin{abstract}
We study the potential of adaptive optics (AO) to protect entanglement of high-dimensional photonic orbital-angular-momentum (OAM) states against turbulence-induced phase distortions. 
We demonstrate that AO is able to reduce crosstalk among the OAM modes and, consequently, the entanglement decay as well as photon losses. A test of the AO-stabilized output state against high-dimensional Bell inequalities shows that the transmitted entanglement allows for secure communication, even in the strong scintillation regime.
\end{abstract}


\section{Introduction}
High-dimensional, discrete quantum systems, also referred to as qudits, present several advantages over simple two-level quantum systems (qubits).
The first and rather obvious benefit of high-dimensional states is their capability of naturally increasing the information encoded in a single carrier.
Moreover, when compared to qubits, qudits are more robust against quantum cloning \cite{CerfPRL2002} and, thus, they allow to enhance the security of quantum key distribution (QKD)\cite{QuantCryptReview}. 
Some promising QKD protocols are based on quantum entanglement \cite{E91, Lo2014}. 
In this case, the presence of an eavesdropper can be detected via Bell inequalities, which are the more violated the larger the dimensionality of the employed entangled states \cite{DadaNatPhysLett2011,PhysRevLett.85.4418}. 
Therefore, entangled qudits can be used to outperform two-dimensional QKD schemes \cite{PhysRevA.88.032309}. 
This makes qudit states potentially useful in free-space quantum communication.

Photonic orbital angular momentum (OAM) states \cite{allen92} are suitable candidates for the realization of such high-dimensional quantum systems. OAM spans a discrete, infinite-dimensional Hilbert space, and photon pairs entangled in this degree of freedom are naturally produced in spontaneous parametric down conversion (SPDC)  \cite{DadaNatPhysLett2011,PhysRevA.65.033823,Mair2001}. 
On the downside, the advantages associated with OAM can be nullified by a turbulent atmosphere, because the defining feature of OAM-carrying light beams, namely their helical wave front, is fragile with respect to turbulence-induced refractive index fluctuations \cite{Paterson2005}. 

The behaviour of OAM-entangled qubit pairs in turbulence has been actively studied theoretically \cite{Smith2006,Gopaul2007,Leonhard2015,PhysRevA.92.012326} and experimentally, both in laboratory simulations \cite{IbrahimPRA2013} and through a $3$~km free-space channel \cite{Krenn02112015}. 
In contrast, the effect of atmospheric turbulence on high-dimensional OAM states remains largely unexplored. 
To the best of our knowledge, the evolution of entangled qutrits has only been studied theoretically \cite{TobiQutrits} and in one laboratory simulation \cite{ZhangPRA2016}. 
Concerning higher-dimensional systems, the only known result is a free-space QKD experiment with $4$-dimensional quantum states (realised via the hybridization of the helical wave front and of the polarisation of the photon), where an increased data rate with respect to two-dimensional encoding was demonstrated over a distance of $300$~m \cite{QKDOttawa}. 
However, to achieve longer propagation distances, compensation schemes for turbulence-induced errors, for example with methods of adaptive optics (AO), are needed.

During the past few decades, AO has developed into an indispensable tool for large, earth-based astronomical telescopes \cite{Milonni1999,tysonAO}. Based on wavefront measurements of the received light, corrective elements such as deformable and tip-tilt mirrors are steered by a real-time control loop to correct for turbulence-induced phase distortions. In addition, AO can stabilize free-space optical communication links which are also distorted by atmospheric turbulence \cite{Levine:98}. 
In recent experiments, the capability of AO to mitigate the crosstalk of a classical OAM-multiplexed beam has been investigated \cite{Ren2014a,Ren2014,Rodenburg2014,Li2014}. 
Furthermore, theory \cite{Leonhard2018} did demonstrate  the potential of AO to reduce signal and entanglement losses of OAM qubits, even under moderate turbulence conditions.

In the present contribution, we investigate the ability of AO to protect maximally entangled states of two qudits of dimension $d \in \{2,3,4\}$ in a turbulent channel, for a broad range of atmospheric conditions. After identifying states that can be efficiently protected, we characterize the entanglement evolution in different dimensions in terms of the respective violations of suitably generalized Bell inequalities \cite{CollinsPRL2002}. On the one hand, our results indicate that stronger violations of Bell inequalities, as achievable for high-dimensional states, cannot compensate for their faster degradation under turbulence. In other words, entanglement fades away more rapidly for higher-dimensional states. On the other hand, we show that AO is capable to significantly increase the maximal turbulence strength at which entanglement vanishes. AO hence bears the potential to render the entanglement distribution of OAM qudits practicable in free-space quantum communication.

The paper is organized as follows: In \sref{Sec:FreeSpace}, we recollect the main features of light propagation in turbulent media and present our numerical simulation routine. 
The capability of AO to reduce the turbulence-induced spreading of the OAM spectra of single photons is discussed in \sref{Sec:single}, while its impact on entanglement evolution is presented in \sref{Sec:Ent}. At last, in \sref{Sec:BellParameter}, we discuss the violation of generalized Bell inequalities. \Sref{Sec:conc} concludes our work.

\section{Description of the free-space channel}
\label{Sec:FreeSpace}
\subsection{Characterizing atmospheric turbulence}
\label{sec:turb}
When an optical wave is transmitted through a free-space atmospheric channel, it experiences diffraction, as well as phase and intensity fluctuations induced by small random fluctuations of the refractive index. 
The latter can be decomposed into an average value 1 and small fluctuations $\delta n({\bi r})$, as $n({\bi r}) \approx 1 + \delta n({\bi r})$. Typically, $\delta n({\bi r})\sim 10^{-3}$ is assumed to be described by a Gaussian random field \cite{andrews}. Within Kolmogorov's theory of turbulence, the statistics of 
the field $\delta n ({\bi r})$ is controlled by its power spectral density 
{$\Phi (\bkappa) = 0.033C_n^2|{\bkappa}|^{-11/3}$}, with ${\bkappa}$ the spatial wave vector, and $C_n^2$ the refractive index structure constant which quantifies the turbulence strength \cite{andrews}.

In the paraxial approximation
\footnote{The latter assumes that the optical wave $\psi({\bi r})$ is a slowly varying function of the propagation direction $z$, such that $|\frac{\partial^2 \psi}{\partial z^2}| \ll| \nabla^2_T\psi|$ and $|\frac{\partial^2 \psi}{\partial z^2}| \ll k|\frac{\partial \psi}{\partial z}|$.},
the (horizontal) transmission of a monochromatic light beam across such medium   obeys the \textit{stochastic parabolic equation} 
\begin{equation}
- 2 i k \frac{\partial}{\partial z} \psi({\bi r}) = \nabla^2_T \psi({\bi r}) +
 2k^2\delta n({\bi r})\psi({\bi r}),
\label{StochPar}
\end{equation}
where $k = 2\pi/\lambda$, $\lambda$ the wavelength, ${\bi r}$ the three-dimensional position vector, $z$ the propagation direction and $\nabla^2_T $ the two-dimensional Laplacian in the transverse plane \cite{andrews}. The first and second terms on the right hand side of \eref{StochPar} describe the wave's diffraction and refraction, respectively, and can be interpreted as kinetic and potential energy terms \cite{Cook:75} which generate the unitary Schr\"odinger evolution of the transverse amplitude pattern $\psi(x,y; z)$ parametrized by the generalized time $z$. The unitarity property  will be a useful fact in sections \ref{Sec:single} and \ref{Sec:Ent} below.
 
Let us now introduce the remaining quantities which define the influence of diffraction and atmospheric turbulence on the propagating wave. The diffraction of a beam with width $w_0$ is characterised by the Rayleigh range $z_R=\pi w_0^2/\lambda$ \cite{andrews}. 
Combinations of $C_n^2$, $\lambda$ (or $k$), and $z$ yield the Fried parameter and the Rytov variance -- two key parameters that capture atmospheric effects. The former sets the typical transverse length scale on which phase errors are correlated, and reads \cite{andrews}, for a plane wave, 
\begin{equation}
r_0 = (0.423k^2C_n^2z)^{-3/5}.
\label{Fried}
\end{equation}
The Rytov variance,
\begin{equation}
\sigma_R^2 = 1.23C_n^2k^{7/6}z^{11/6},
\end{equation}
is a measure of the strength of the intensity fluctuations (scintillation) which arise when phase perturbations are combined with subsequent propagation. Scintillation is considered to be weak if $\sigma_R^2 < 1$, and strong otherwise.  
 
It is convenient to introduce two dimensionless quantities, the renormalized propagation distance $t = z/z_R$ \cite{PhysRevA.90.052115} and the turbulence strength $\mathcal{W} = w_0/r_0$ 
\cite{Paterson2005,PhysRevA.90.052115,Paterson2004}. 
Indeed, by fixing $\mathcal{W}$ and $t$, one can deduce  the Rytov variance $\sigma_R^2 = 1.63\mathcal{W}^{5/3}t^{5/6}$ and, thus, define the scintillation regime. Knowledge of the Rytov variance is important for the implementation of our numerical method, which we present below in {section~}\ref{Sec:num}. Also note that the dimensionless parameters $t$ and $\mathcal{W}$ are combined from the beam and turbulence parameters. Therefore, a rescaling of the latter which keeps $t$ and $\mathcal{W}$ unaffected allows one to infer results for different atmospheric conditions and beam geometries without calculation. We provide an example thereof in the following section.

\subsection{Numerical method}
\label{Sec:num}

We solved \eref{StochPar} using the {\it split-step} method \cite{lukin,schmidt}, which is a versatile numerical tool to model a broad range of turbulence conditions. The solutions obtained with this approach were previously verified experimentally \cite{lukin} for classical beam propagation. Furthermore, the split-step method provides good quantitative agreement with analytical solutions for the entanglement evolution of OAM qubit states in turbulence \cite{Leonhard2018,PhysRevA.90.052115}.
 
The main idea of the method is to replace the three-dimensional turbulent medium between the transmitter and the receiver by a sequence of equidistant thin screens. The screens mimic the refraction of a wave on $\delta n({\bi r})$ (random phase errors), whereas between the screens the wave undergoes free diffraction in vacuum.
In our particular implementation, the random phase screens were generated using the subharmonic method presented in \cite{Lane1992}, while the free diffraction in vacuum was performed with an angular spectrum propagator \cite{goodman, schmidt}.

The number of  screens must be chosen such that at each partial propagation step, consisting of two vacuum propagations and a single phase screen in the middle, the weak scintillation condition  ($\sigma_R^2 <1$) is satisfied.  To ensure this latter constraint, we imposed an even stricter condition, $\sigma_R^2 < 0.5$.
In our numerical simulation, we fixed the total renormalized propagation distance to $t=0.19$, and considered $20$ values of the turbulence strength $\mathcal{W} \in [0; 4.9 ]$. 
This corresponds to a Rytov variance $\sigma_R^2\leq 5.8$ for the total path. 
Then the requirement $\sigma_R^2 < 0.5$ for the partial propagation steps was accomplished by splitting the full distance into $21$ steps.  
Finally, to mimic realistic detection conditions, we assumed a collecting aperture with a finite diameter in the receiver plane. 
In addition, in order to reduce losses caused by finite size effects, the collecting aperture was chosen such that a beam twice as large as the OAM modes arriving at the receiver would fit into it.

Let us finally illustrate the rescaling mentioned at the end of \sref{sec:turb} above. 
The values $t=0.19$ and $\mathcal{W}=4.9$ can be derived, for instance, from a refractive index structure constant $C_n^2 = 6.67 \times 10^{-13}$~m$^{-2/3}$ (moderate turbulence), wavelength $\lambda = 1064$~nm, beam waist $w_0=0.03$~m, and propagation distance $z = 500$~m, which are the same values as in  \cite{Leonhard2018} (apart from the slightly weaker turbulence there). But the same $t$ and $\mathcal{W}$ can also be obtained for $C_n^2 = 2.9 \times 10^{-14}$~m$^{-2/3}$, same $\lambda$, $w_0=0.0735$~m, and $z = 3000$~m. 
In fact, we chose the latter values of $\lambda, w_0$ and $z$  in our simulations, because they are close to the experimental parameter settings used in \cite{Krenn02112015}.

\subsection{Adaptive Optics}
\label{Sec:AO}
An AO system can correct for turbulence-induced phase distortions by first measuring the phase and then computing the optimum correction which is typically introduced by deformable and tip-tilt mirrors.  
An auxiliary light beam, the beacon, senses the wavefront distortions, and is separated  from the OAM signal by use of, e.g., a distinct polarization.  We assume that the beacon with a Gaussian transverse intensity profile  propagates prior to and along the same channel as the OAM photons \cite{Leonhard2018}. 
The measured phase  $\varphi_B(\brho)$ is then used to correct the photon's wavefront, which leads to the corrected OAM state ${\tilde{\psi}(\brho,z) = \exp[-i\varphi_B(\brho)]\psi(\brho,z)}$, with $\brho$ the coordinate in the transverse plane. 

We compare two different AO models -- which we introduced in \cite{Leonhard2018} -- with simulations of the uncorrected photons. The first AO model  consists  in essentially a full reconstruction of the phase profile of the propagated beacon, with a resolution only limited by the grid underlying the computation. 
This represents the ideal AO, actually unachievable in a real experiment, since no limitation of a real AO system is accounted for.
In contrast, the second AO model only stabilizes the direction of the incoming light beam, with a tip-tilt mirror. 
Such a mirror can merely rotate in the receiver plane and, thus, only corrects for the two lowest-order modes (tip and tilt) of atmospheric turbulence, which, however, are dominant in the weak turbulence regime \cite{noll:76}.  
In a real experiment, a position-sensitive detector or a four quadrant diode is placed in the focal plane of a lens which, by its Fourier-transforming properties, converts a tilted input wavefront into a displaced focal spot \cite{tysonAO}. 
The required mirror tilt follows from the spot displacement and the focal length of the lens. 
Hence, the tip-tilt compensation represents the simplest version of AO, which is typically outperformed by any real AO system. 

\section{Evolution of single OAM modes}
\label{Sec:single}

We wish to describe the free-space transmission of photonic states entangled in their OAM degree of freedom. 
The main source of signal and entanglement losses for this kind of states is the turbulence-induced coupling (crosstalk) between different OAM modes \cite{Smith2006, Gopaul2007, Leonhard2015, PhysRevA.92.012326,IbrahimPRA2013}. 
To gain some insight into this effect and its partial reversal by AO, we set out to investigate the evolution of a single OAM mode under turbulence, given different levels of AO-aided correction.

We study single photons populating Laguerre-Gaussian (LG) modes $u_{0,l}(\rho,\theta,z)$, i.e.,  with radial index $p=0$ and azimuthal index $l$:
\begin{eqnarray*}
u_{0,l}(\rho,\theta,z) =& \frac{\mathcal{A}_{p,l}}{\sqrt{w(z)}}\left(\frac{\sqrt{2}\rho}{w(z)}\right)^{|l|}\exp\left(-\frac{\rho^2}{w^2(z)}\right)\exp\left[-ik\frac{\rho^2z}{2(z^2 +z_ R^2)}\right]\\
&\times \exp(i l \theta)\exp\left[-i(|l| +1)\Phi(z)\right],\nonumber
\end{eqnarray*}
where $(\rho,\theta)$ are the polar coordinates in the transverse plane, $w(z) = w_0\sqrt{1+(z/z_R)^2}$ is the $z$-dependent beam waist, $\Phi(z) = \arctan(z/z_R)$ is the Gouy phase and $\mathcal{A}_{p,l}$ is a normalization constant, the explicit form of which is not relevant here \cite{AndrewsBabiker}. 
We use the shorthand notation $\ket{l}$ for such modes.

The evolution of an input mode $\ket{l_0}$ under the dynamics described by \eref{StochPar}, for a particular realization $i$ of the turbulent fluctuations of the refractive index, is described by a unitary operator $\hat{U}_{turb}^{(i)}(z)$, which transforms the input state into a superposition of many OAM modes,
\begin{equation}
\ket{\psi^{(i)}_{l_0}} = \hat{U}_{turb}^{(i)}(z)\ket{l_0} = \sum_{l}c^{(i)}_{l,l_0}(
z)\ket{l},
\label{unitary}
\end{equation}
with $c^{(i)}_{l,l_0} = \bra{l}\hat{U}_{turb}^{(i)}\ket{l_0}$. 
The action of AO described in Sec.~\ref{Sec:AO} can be expressed by a unitary operator $\hat{U}_{AO}:=\exp[-i\varphi(\brho)]$. Thus, the AO-corrected version of \eref{unitary} is obtained by replacing $\ket{\psi^{(i)}_{l_0}}$ with $\ket{\tilde{\psi}^{(i)}_{l_0}} = \hat{U}_{AO} \ket{\psi^{(i)}_{l_0}}$. 
\begin{figure}
\includegraphics[width=\textwidth]{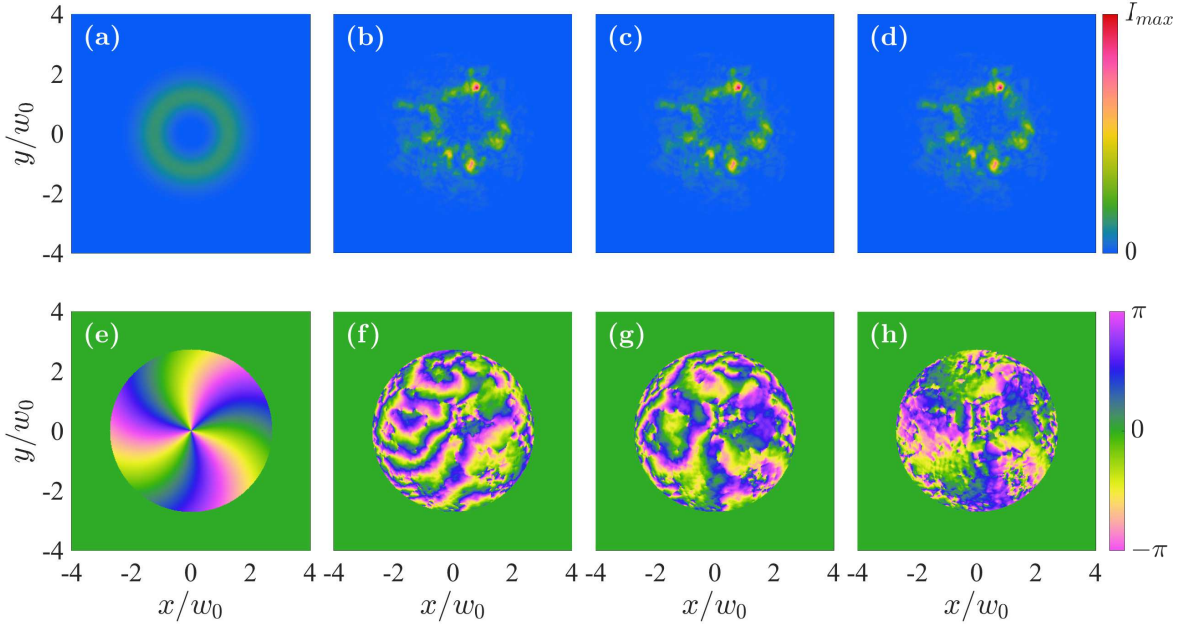}
\caption{Intensity (a-d) and phase (e-h) distributions for an LG mode with azimuthal index $l_0 = 3$, for a single realization of atmospheric turbulence: (a, e) mode propagated in absence of turbulence; (b, f) turbulence-distorted mode; (c, g) tip-tilt corrected and (d, h) ideally corrected mode. The specific turbulence realization used for this plot corresponds to $\mathcal{W}= w_0/r_0 = 2.45$. Note that the color scales for all intensity (a-d) and phase plots (e-h) are the the same.}
\label{Fig:IntensityPhase}
\end{figure}

In \fref{Fig:IntensityPhase}(b, f), the intensity and phase distributions associated with the state $\ket{\psi^{(i)}_{l_0=3}}$ are presented, and compared with those of the mode propagated in absence of turbulence (a, e), as well as the tip-tilt corrected (c, g) and the ideally corrected (d, h) modes. 
The transmitted states are monitored only within a finite circular aperture in the receiver plane, which is evident from the phase plots, but not noticeable in the intensity distributions which smoothly decay to zero within the aperture  -- indicating that the latter was chosen large enough to avoid aperture-induced losses.

Inspection of \fref{Fig:IntensityPhase}(b) shows that the intensity distribution of the output state still resembles the doughnut shape of the unperturbed mode in \fref{Fig:IntensityPhase}(a), garnished by turbulence-induced distortions. 
Furthermore, the intensity patterns in figures \ref{Fig:IntensityPhase}(b-d) are identical, which immediately follows from the fact that AO only performs phase corrections. In contrast, the associated phase distributions in figures \ref{Fig:IntensityPhase}(f-h) are rather different from each other and from the unperturbed phase distribution shown in \fref{Fig:IntensityPhase}(e). \Fref{Fig:IntensityPhase}(f) shows that the input phase in \fref{Fig:IntensityPhase}(e) is largely obliterated by turbulence. 
It is noteworthy that, despite some residual phase errors, the ideal AO correction produces a phase distribution [\fref{Fig:IntensityPhase}(h)] that clearly mirrors the one of the mode propagated in absence of turbulence [\fref{Fig:IntensityPhase}(e)]. 
When only the tip-tilt correction is applied [\fref{Fig:IntensityPhase}(g)], even though the phase distribution maintains a distorted structure, the length-scale of phase variations is larger than in the uncorrected state. 
This qualitative analysis of intensity and phase distributions distorted by a particular realization of turbulence already shows the main limitations of AO: intensity distortions are not corrected and residual phase errors persist even in the case of an ideal correction.
These residual phase errors are most likely caused by the different beam profiles of the Gaussian beacon and the OAM photon.

\begin{figure}
\includegraphics[width=\textwidth]{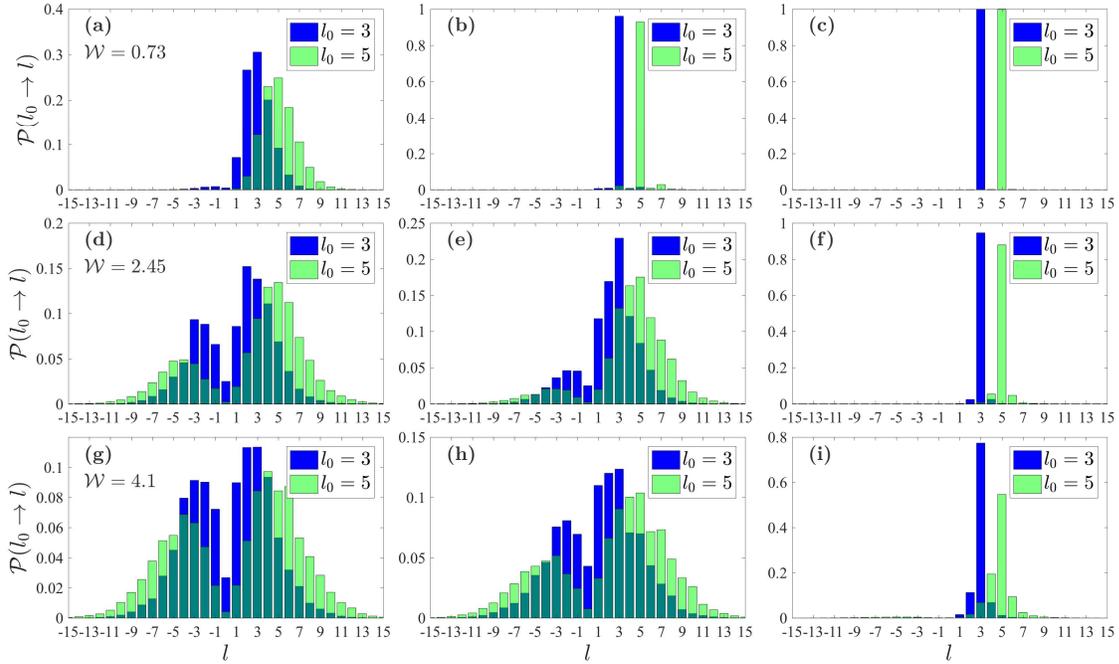}
\caption{Spiral spectra for turbulence-distorted LG modes with $l_0 = 3,5$, averaged over $N=500$ realizations of turbulence, for different degrees of correction [(a, d, g) no correction, (b, e, h) tip-tilt  and (c, f, i) ideal correction] and turbulence strengths [(a-c) $\mathcal{W} = w_0/r_0 = 0.73$ , (d-f) $\mathcal{W} = 2.45$  and (g-i) $\mathcal{W} = 4.1$]. Note the different scales of the $\mathcal{P}-$axes, in the different panels}
\label{Fig:Spectra}
\end{figure}

For a more quantitative assessment of the efficiency of AO, we now study the OAM distribution of the turbulence-distorted, ideally and tip-tilt corrected states. 
From the mode expansion in \eref{unitary}, we can extract the probability of an input mode $l_0$ to scatter into another OAM mode $l$,
\begin{equation}
\mathcal{P}(l_0\to l) = \frac{1}{N}\sum_{i=1}^N |c^{(i)}_{l,l_0}|^2,
\label{spectra}
\end{equation}
where the sum over $i$ in \eref{spectra} represents the disorder average over $N$ realizations of atmospheric turbulence
\footnote{Different realizations of the turbulent channel are obtained using different draws of the 21 random phase screens used in the numerical simulation routine described in \sref{sec:turb}}. 
The quantity defined in \eref{spectra} is also known as the {\it spiral spectrum} \cite{DiLorenzoPeresPRL2010}. 

\Fref{Fig:Spectra} shows the spiral spectra of the states $\ket{\psi_{l_0}}$ for $l_0 = 3,5$ averaged over $N = 500$ realizations of turbulence, each for three different values of the turbulence strength $\mathcal{W}$, such that the central row of \fref{Fig:Spectra} corresponds to the same turbulence strength as in \fref{Fig:IntensityPhase}. In \fref{Fig:Spectra} (d), the spiral spectra are composed of two peaks, a higher one 
around the input mode and a smaller one, located almost symmetrically with respect to the $l = 0$ mode. This double-peak distribution is due to the fact that the phase information is severely distorted, while the doughnut structure of the intensity is nearly preserved, leading to a reduced overlap with the fundamental Gaussian mode \cite{Anguita:08}. For weak turbulence [\fref{Fig:Spectra}(a)], the peak resulting from crosstalk is suppressed, whereas for stronger turbulence  [\fref{Fig:Spectra}(g)] the heights of the two peaks in the spiral spectra tend to equalize (note the different scales of the $\mathcal{P}-$axes, in \fref{Fig:Spectra}), featuring a loss of the encoded information. 

In the rightmost column of \fref{Fig:Spectra}, we see that ideal AO almost restores the input spectra, and small residual populations only affect modes adjacent to the input one. Tip-tilt AO (central column of \fref{Fig:Spectra}) only partly reduces the crosstalk peak and enhances the input mode contribution.
Comparison of the $l_0 = 3$ (blue) and $l_0 = 5$ (green) results shows that the spiral spectra become broader for larger OAM, whereas AO becomes less effective.

\section{Evolution of maximally entangled states}
\label{Sec:Ent}
\subsection{Model}
The physical intuition gained from our study of single 
photon states above is now used to understand the decay of high-dimensional entanglement in the atmosphere, and to assess the potential of AO to prevent such decay. Let us first describe the specific scenario we have in mind. Let Alice prepare the maximally entangled state
\begin{equation}
\ket{\psi_0} = \frac{1}{\sqrt{d}}\sum_{l_0\in\mathcal{B}}\ket{l_0,-l_0},
\label{MaxEntState}
\end{equation}
in her lab, where $\mathcal{B}$ represents a $d$-dimensional subspace of the total OAM Hilbert space, and will be referred to as the encoding subspace in the following. 
Subsequently, one of the photons is sent to Bob, through a free-space channel of length $z$. 
For a particular realization $i$ of the turbulent channel, this photon evolves under the action of $\hat{U}^{(i)}_{turb}$, as described in the previous paragraph. 
In the output plane in Bob's lab, observables are only measured within the encoding subspace.  
The combined effect of propagation across the turbulent medium and projection onto the encoding subspace can be represented as the completely positive (non-trace preserving) map 
$\hat{\mathcal{K}}^{(i)} = \hat{\Pi}_\mathcal{B} \hat{U}^{(i)}_{turb}$, with $\hat{\Pi}_\mathcal{B}$ the projector onto the encoding subspace
\footnote{Yet another source of non-unitarity is the finite size aperture in the receiver plane. 
We however assume that this effect can be absorbed in $\hat{\Pi}_\mathcal{B}$, and will remain subdominant as long as the receiver aperture is chosen large enough.}. 
The output biphoton state is then given by
\begin{equation}
\ket{\psi^{(i)}} = \left[ \mathbb{1}\otimes\hat{\mathcal{K}}^{(i)}\right]\ket{\psi_0
}=\sum_{l_0,l\in{\cal B}} \frac{c^{(i)}_{l_0,l}}{\sqrt{d}}\ket{l_0,l},
\label{OutState}
\end{equation}
where we act with the identity operator $\mathbb{1}$ on the photon remaining in Alice's lab.
Upon disorder average, the pure state \eref{OutState} turns into a mixed state described by the density matrix
\begin{equation}
\hat{\rho} = \frac{1}{\mathcal{T}}\sum_{i=1}^N \ket{\psi^{(i)}}\bra{\psi^{(i)}},
\label{DM}
\end{equation}
where 
\begin{equation}
\mathcal{T} = \Tr\left[\sum_{i=1}^N \ket{\psi^{(i)}}\bra{\psi^{(i)}}\right]
\label{trace}
\end{equation}
renormalizes the state after the non-unitary projection.

To quantify the entanglement of the disorder-averaged output density matrix in the qubit case ($d=2$), we use concurrence \cite{Wootters1998}
\begin{equation}
 C(\hat{\rho}) = \mathrm{max} \{ \sqrt{\lambda_1} -\sqrt{\lambda_2} - \sqrt{\lambda_
3} -\sqrt{\lambda_4},0 \},
 \label{concurrence}
\end{equation}
where the $\lambda_i$ are the eigenvalues, in decreasing order, of the matrix $\hat{\mathcal{R}} = \hat{\rho} (\hat{\sigma}_y\otimes\hat{\sigma}_y)\hat{\rho}^* (\hat{\sigma}_y\otimes\hat{\sigma}_y)$, with $\hat{\sigma}_y$ the second Pauli matrix. 
For $d>2$, we employ negativity \cite{Zyczkowski1998,Vidal2002} 
\begin{equation}
\mathcal{N}(\hat{\rho}) = \frac{||\hat{\rho}^{PT}||_1 - 1}{2 \mathcal{N}_{max}},
\label{Negativity}
\end{equation}
where $||\hat{\rho}^{PT}||_1$ denotes the trace norm of the partially transposed density matrix $\hat{\rho}^{PT}$. 
We divide the usual definition of negativity by its 
dimension-dependent maximum $\mathcal{N}_{max} = (d -1)/2$, for a better comparison of entanglement evolution in different dimensions. 
Recall that negativity has the advantage of being easily computable for mixed states in arbitrary dimensions, but it fails to recognize positive-partial-transpose (PPT) states \cite{GuhneToth2009}. 

Before proceeding with the results of our numerical simulation, we note that in an encoding subspace $\mathcal{B}$ of dimension $d$, there are $d^2$ orthogonal generalized Bell states \cite{BennetPRL1993}. 
However, we consider the evolution through a one-sided channel (see \eref{OutState}) and, as a consequence of the Choi-Jamio\l{}koswki isomorphism \cite{Jamiolkowski1972,Choi1975}, all maximally entangled states of dimension $d$ experience the same entanglement evolution. 
Thus, we can fully characterize the behaviour of all maximally entangled states by considering a single state \eref{MaxEntState} for each choice of the encoding subspace $\mathcal{B}$.

\subsection{Results for qubit states}
\label{Sec:qubits}

Let us start by exploring the capability of adaptive optics to enhance entanglement transmission through turbulence when qubit encoding is employed: We consider an encoding subspace spanned by two modes with opposite OAM ($\mathcal{B} = \{-l_0,l_0\}$, with $l_0\leq5$). 
We already studied this scenario in \cite{Leonhard2018}, for the same renormalized propagation distance $t = 0.19$, but for a narrower range of turbulence strengths $\mathcal{W}<2$, using four phase screens. 

\begin{figure}
\includegraphics[width=\textwidth]{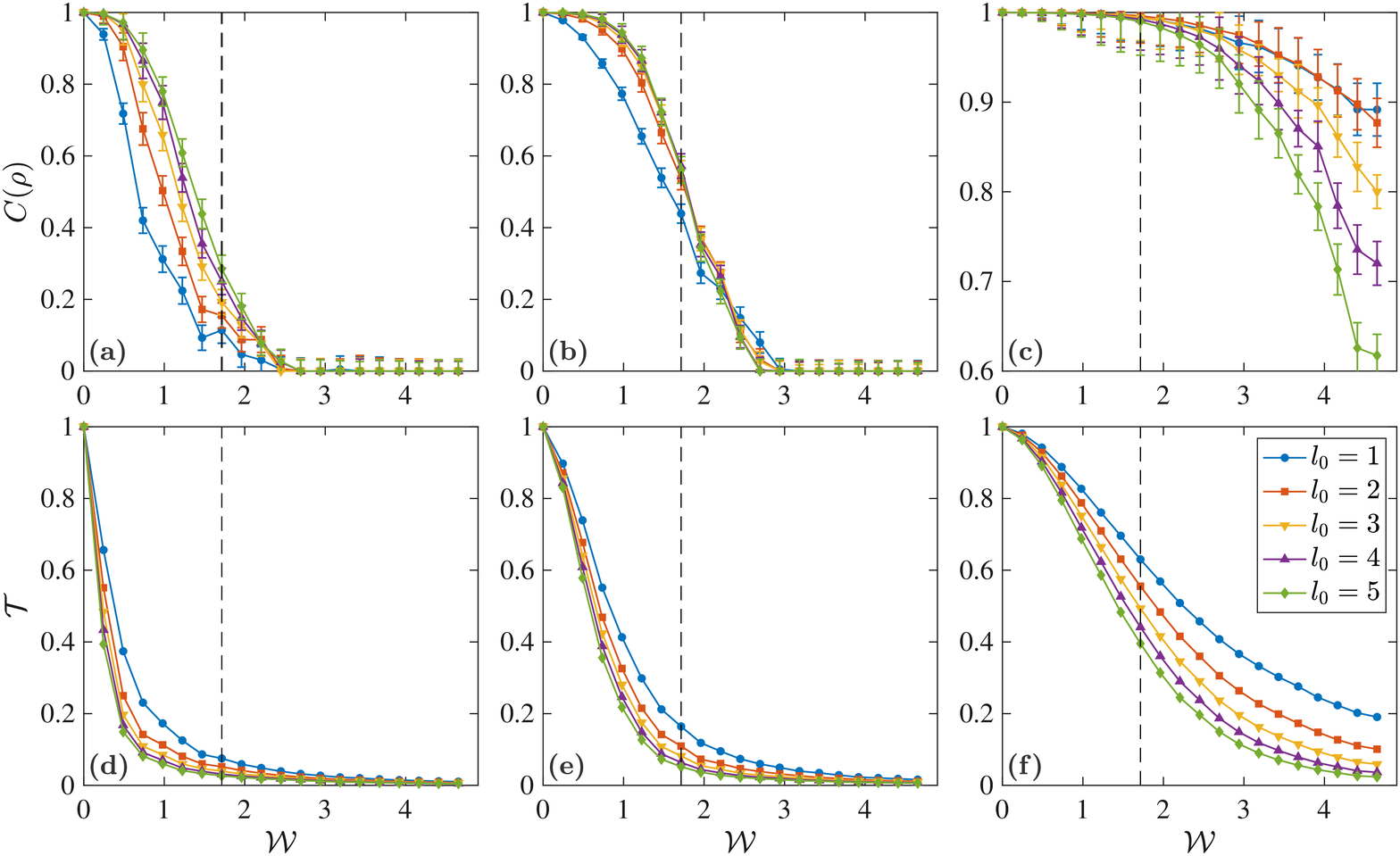}
\caption{Concurrence $C$ (a-c) according to \eref{concurrence} and trace $\mathcal{T}$ [see \eref{trace}] (d-f) of the disorder-averaged output density matrix $\rho$ as a function of the disorder strength $\mathcal{W}=w_0/r_0$ for $\mathcal{B} = \{-l_0,l_0\}$ (qubit encoding). 
Different degrees of AO correction are considered: (a,d) no, (b, e) tip-tilt and  (c, f) ideal correction. Note the different scale of the $C-$axis in panel (c).
The dashed vertical lines at $\mathcal{W} \approx 1.7$ correspond to a Rytov variance $\sigma_R^2 = 1$, indicating the transition from weak to strong scintillation. Error bars of concurrence in (a-c) are inferred via error propagation from the standard deviations around the mean value of the density matrix elements, calculated from $N=500$ realizations of turbulence.
Standard deviations in (d-f) are smaller than the symbol size.}
\label{Fig:EntanglementQubits}
\end{figure}

We thus expect to find the same entanglement evolution for turbulence strengths $\mathcal{W}<2$ in our present model with 21 phase screens.
Indeed, this is our observation in the top row of \fref{Fig:EntanglementQubits}, where the concurrence evolution in turbulence is shown (a) without, (b) with tip-tilt and (c) with ideal AO corrections. 
Interestingly, the transition from weak to strong scintillation ($\sigma_R^2 = 1$) denoted by the dashed vertical line in \fref{Fig:EntanglementQubits} occurs at $\mathcal{W} \approx 1.7$, which was close to the boundary of the turbulence range considered in our previous work \cite{Leonhard2018}, and thus could not be clearly identified.

In the strong scintillation regime, intensity fluctuations, which cannot be corrected for by AO, become more and more pronounced (see also figures \ref{Fig:IntensityPhase} and \ref{Fig:Spectra}). 
Therefore, it is not surprising that AO correction becomes less effective in this regime.  
In particular, for ideally corrected states [\fref{Fig:EntanglementQubits}(c)], we witness a significantly enhanced entanglement decay once the threshold identified by $\sigma_R^2 = 1$ is crossed. 
Hence, we conclude that scintillation imposes the fundamental limit on AO efficiency.
The enhanced decay behavior beyond $\sigma_R^2 = 1$ also reveals a better stability of lower OAM modes which we have not noticed in our earlier work \cite{Leonhard2018}.
This is in contrast to the trend observed in the uncorrected case [\fref{Fig:EntanglementQubits}(a)]. Since turbulence-induced phase distortions depend on the input intensity and phase profiles, the difference in geometry between the OAM modes and the Gaussian beacon beam may cause the reduced AO efficiency for larger azimuthal indices, and thus the opposite $l_0$ dependence in figures \ref{Fig:EntanglementQubits}(a) and (c). In \fref{Fig:EntanglementQubits}(b), all curves, apart from the one for$l_0 = 1$, collapse onto one line. 
In contrast to the ideally corrected case, the speed-up in the entanglement decay starts earlier than at the $\sigma_R^2 = 1$ line (roughly, at half the corresponding turbulence strength), where other turbulence-induced phase distortions (e.g. astigmatism, coma, spherical aberration \cite{noll:76}) become relevant. Furthermore, the transition to the strong scintillation regime does not strongly affect the slope of the decay curve in [\fref{Fig:EntanglementQubits}(b)]. 
This means that higher-order phase aberrations can be as detrimental for entanglement transmission as scintillation-induced intensity fluctuations. 

The bottom row of \fref{Fig:EntanglementQubits} presents the decay of the trace $\mathcal{T}$ of the disorder-averaged output density matrix as a consequence of turbulence-induced scattering into modes outside the encoding subspace $\mathcal{B}$.
We see that both tip-tilt [\fref{Fig:EntanglementQubits}(e)] and ideal [\fref{Fig:EntanglementQubits}(f)] corrections result in a prominent trace enhancement, as compared to the uncompensated case [\fref{Fig:EntanglementQubits}(d)].
For example, for $\mathcal{W} = 1.96$, tip-tilt AO increases the trace by a factor of two, whereas ideal AO achieves an enhancement of an entire order of magnitude. 
Trace enhancement implies a larger number of collected photons, and is thus beneficial for an improved signal-to-noise ratio. 

Let us conclude this section with a discussion of the different AO efficiencies with respect to the output state's entanglement as compared to its trace.
As mentioned in section \ref{Sec:single}, turbulence-induced intensity and phase fluctuations lead to a spreading of the spiral spectrum. 
This indicates intermodal scattering not only  inside the encoding subspace $\mathcal{B}$, but also beyond it. 
From the third column in \fref{Fig:Spectra}, we see that (at $\mathcal{W} = 2.45$) crosstalk between the modes $+ l_0$ and $- l_0$ is indiscernible. Consequently, entanglement reduction is weak. 
On the contrary, non-negligible populations of several modes outside the encoding subspace (especially, of adjacent modes $l_0 \pm 1$) signify a relatively large trace decay, which implies a stronger effect of turbulence-induced errors on the output state's trace than on its entanglement, as consistently observed by comparison of figures \ref{Fig:EntanglementQubits}(c) and (f).

\subsection{Results for qutrit states}
\begin{figure}
\includegraphics[width=\textwidth]{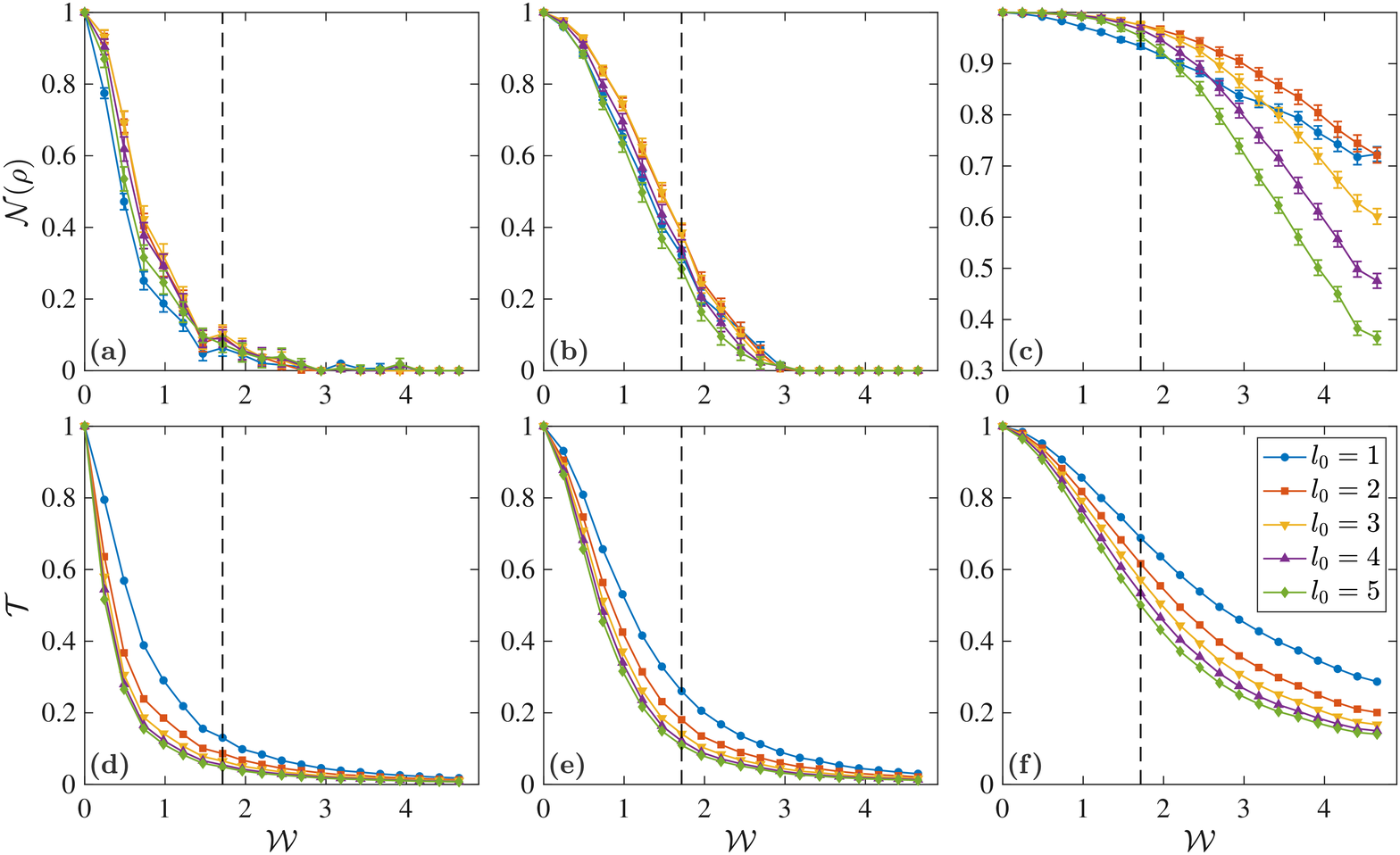}
\caption{Negativity $\mathcal{N}$ according to \eref{Negativity} (a-c) and trace $\mathcal{T}$ [see \eref{trace}] (d-f)  of the disorder-averaged output density matrix $\rho$, plotted against the turbulence strength $\mathcal{W}=w_0/r_0$ for $\mathcal{B} = \{-l_0,0,l_0\}$ (qutrit encoding).
Different degrees of AO correction are considered: (a, d) no, (b, e) tip-tilt and (c, f) ideal corrections. Note the different scale of the $\mathcal{N}-$axis in panel (c).
The dashed vertical lines indicate a Rytov variance $\sigma_R^2 = 1$.
Error bars of negativity in (a-f) are inferred via error propagation from the standard deviation around the mean value of the density matrix elements, calculated from $N=500$ realizations of turbulence. In (d - f), the error bars are smaller than the symbols.
}
\label{Fig:EntanglementQutrits}
\end{figure}

After the discussion of OAM qubits, we now increase the capacity of our encoding states by adding the fundamental 
Gaussian mode to the encoding subspace, i.e. $\mathcal{B} = \{-l_0, 0, l_0\}$, with $l_0\leq5$. 

The top row of \fref{Fig:EntanglementQutrits} shows the negativity evolution under turbulence, (a) without, (b) with tip-tilt and (c) with ideal  corrections. 
The first thing to notice is that, in \fref{Fig:EntanglementQutrits}(a), there is no significant difference between the results for different values of $l_0$. 
A similar result was obtained in the weak scintillation limit, both, theoretically and experimentally, in \cite{ZhangPRA2016}. 
In \fref{Fig:EntanglementQutrits}(c), we can observe the same trend that we saw in the qubit case in the previous section \ref{Sec:qubits}: 
in the weak scintillation regime, the ideal correction proves to be very effective, while in the strong scintillation regime the entanglement decay is enhanced, and more so for larger azimuthal indices. 
The main difference with respect to the qubit case is the behaviour of the $l_0=1$ curve [blue dots in  \fref{Fig:EntanglementQutrits}(c)]. 
It lies below all other curves for $\sigma_R^2 <1$, but crosses them in the strong scintillation region:  
the $l_0 = 5$ curve (green diamonds) at $\mathcal{W} = w_0/r_0 \approx 2$,  the $l_0 = 4$ curve (purple pyramids) at $\mathcal{W} \approx 2.7$, the $l_0 = 3$ curve (yellow triangles) at $\mathcal{W} \approx 3.4$, and the $l_0 = 2$ curve (red squares) at $\mathcal{W} \approx 4.6$. 
Recalling \fref{Fig:Spectra} (f) and (i), the impossibility of the ideal phase correction to reverse scintillation and to fully compensate for the phase errors did result in residual populations of the neighbouring modes. 
Since in our present scenario, the encoding subspace is composed of three adjacent modes for $l_0 =1$, this results in stronger crosstalk in this respective case, and, thus, in a faster negativity decay in the weak scintillation regime. 
For stronger scintillation, instead ideal AO is more efficient in reducing the spreading of the spiral spectra for smaller input OAM modes [see \fref{Fig:Spectra} (c, f, i)] such that the $l_0 =1$ curve gradually overtakes all the others.
In the case of tip-tilt correction, in \fref{Fig:EntanglementQutrits}(b), the result is very similar to the qubit case, where all curves merge with each other.

The bottom row of \fref{Fig:EntanglementQutrits} shows the decay of the trace $\mathcal{T}$ of the disorder-averaged output density matrix. 
The stability  of the fundamental Gaussian mode in turbulence \cite{Anguita:08} decelerates the decay of the trace of the considered qutrit states as compared to the qubit case. 
The trace enhancements obtained with (f) ideal and (e) tip-tilt AO have comparable magnitude as those observed for qubits in \fref{Fig:EntanglementQubits}. 

\subsection{Results for ququart states}
\begin{figure}
\includegraphics[width=\textwidth]{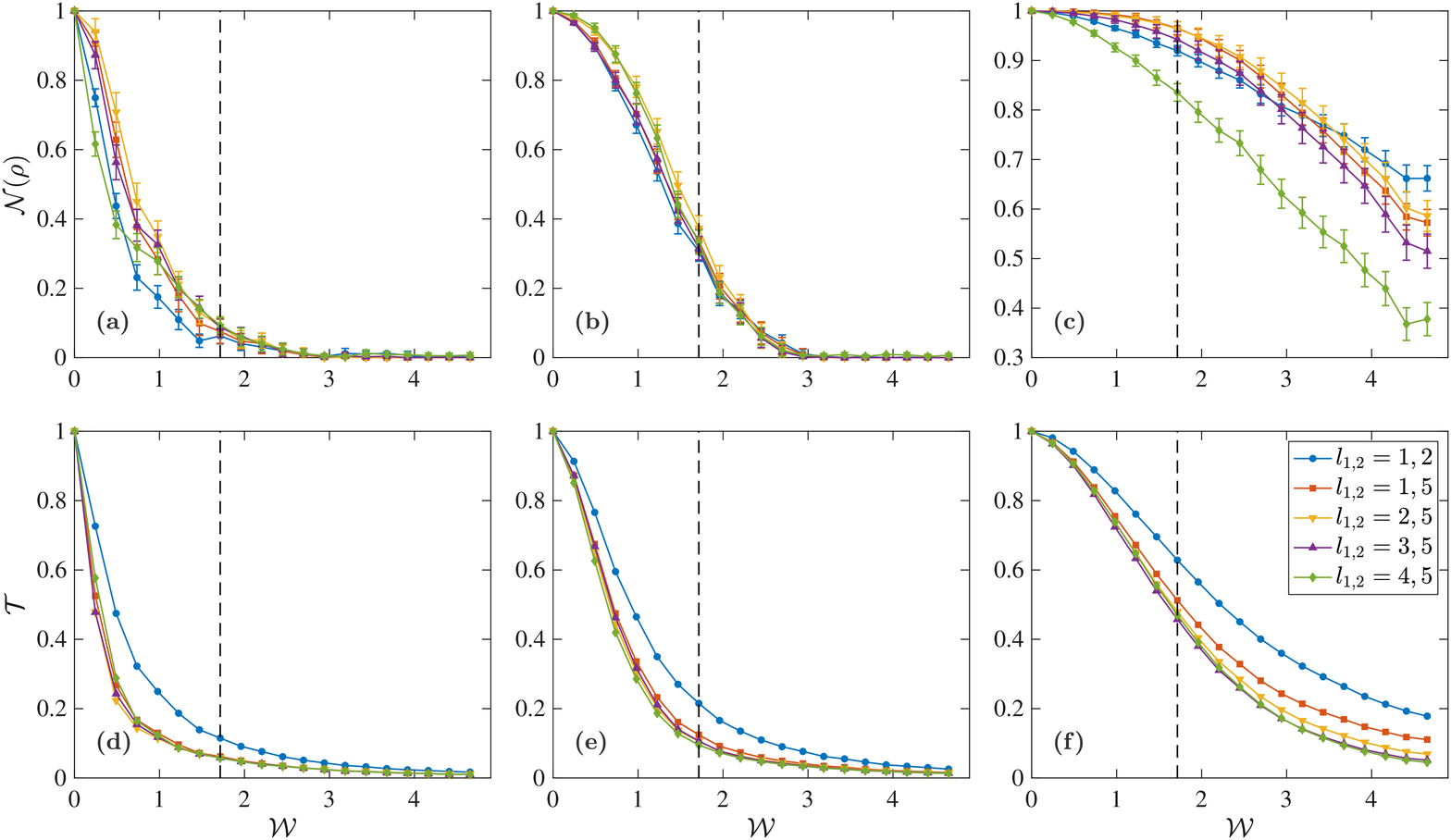}
\caption{Negativity $\mathcal{N}$ according to \eref{Negativity} (a-c) and trace $\mathcal{T}$ [see \eref{trace}] (d-f)  of the disorder-averaged output density matrix $\rho$, plotted against the turbulence strength $\mathcal{W}=w_0/r_0$ for $\mathcal{B} = \{-l_1,l_1,-l_2,l_2\}$ (ququart encoding).
Different degrees of AO correction are considered: (a, d) no, (b, e) tip-tilt and (c, f) ideal corrections. 
The dashed vertical line indicate a Rytov variance $\sigma_R^2 = 1$.
Error bars of negativity in (a-f) are inferred via error propagation from the standard deviation around the mean value of the density matrix elements, calculated from $N=500$ realizations of turbulence. In (d-f), the error bars are smaller than the symbols.
}
\label{Fig:EntanglementQuquarts}
\end{figure}

Finally, we study the entanglement evolution of ququart states under turbulence. 
In particular, we consider $4$-dimensional encoding subspaces composed of two pairs of modes with opposite OAM, i.e. 
$ \mathcal{B} = \{-l_2,-l_1,l_1,l_2 \}$, with  $l_1<l_2\leq 5$. The negativity and trace evolution for five of these ten states is shown in \fref{Fig:EntanglementQuquarts}.

As already observed in the qutrit case, when no correction is applied [\fref{Fig:EntanglementQuquarts}(a)], as well as in the case of tip-tilt correction [\fref{Fig:EntanglementQuquarts}(b)], the entanglement evolution is almost independent of $l_1$ and $l_2$.  
In contrast, for ideal AO the efficiency is sensitive to the set of azimuthal modes included in the encoding subspace [see \fref{Fig:EntanglementQuquarts}(c)]:
in general, larger OAM in the encoding subspace leads to less effective correction. 
For weak scintillation, when encoding into neighbouring modes ($l_2 = l_1 +1$), we observe a lower AO efficiency with respect to the $l_2 > l_1 +1$ case. 
The opposite can be observed for strong scintillation. 
We have already observed this behaviour in the qutrit case and the same explanation applies.

Also the trace decay shown in the bottom row in \fref{Fig:EntanglementQuquarts} presents the same tendency as observed both for qubits and qutrits: independently of the degree of AO correction, higher OAM modes feature a faster decay. 
In the uncorrected case (d), as well as when corrections are applied (e - f), the increase of the encoding subspace induces a slightly improved stability of the trace of ququart states as compared to qubit states. 
However, this effect is weaker than the trace enhancement
observed in the qutrit case by addition of the fundamental Gaussian mode. 

\section{Violation of Bell inequalities and security of quantum key distribution}
\label{Sec:BellParameter}

We conclude this work with a discussion of how entanglement transmission and AO efficiency are affected by the OAM Hilbert space dimension $d$. 
To achieve this goal, we consider a family of Bell parameters $S_d$ \cite{CollinsPRL2002}, which, in case of local correlations, satisfy the inequality 
\begin{equation}
S_d \leq 2,\;\mathrm{for\; all} \; d\geq2.
\label{BellInequality}
\end{equation}
Violation of \eref{BellInequality} not only signals the presence of non-local correlations (entanglement) -- as exploited experimentally to verify high-dimensional OAM entanglement in \cite{DadaNatPhysLett2011}-- but also represents a security check for entanglement-based QKD \cite{E91, GroblacherNJP2006}. 
A Bell parameter can be expressed as the expectation value of a quantum mechanical operator $\hat{S}_d$ and, thus,  computed from the output density matrix of our numerical simulations, as $S_d = \Tr(\hat{S}_d \hat{\rho})$ \cite{BraunsteinPRL1992}. 
The explicit form of the Bell operator $\hat{S}_d$ is given in \ref{BellOperator}. 

\begin{figure}
\includegraphics[width=\textwidth]{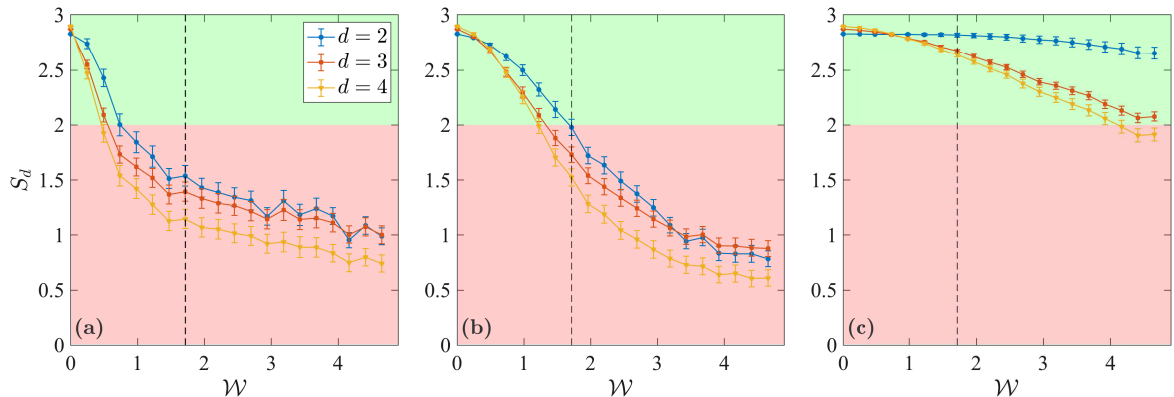}
\caption{Bell parameter $S_d$ [see \eref{Sd}] plotted against $\mathcal{W}$, for different degrees of AO correction: (a) no, (b) tip-tilt and  (c) ideal correction.  The  (red) green shaded area indicates the region where the inequalities \eref{BellInequality} are (not) violated. The dashed vertical lines indicates the disorder strength which defines a Rytov variance $\sigma_R^2 = 1$, i.e. the demarcation line between weak and strong scintillation. Errors are inferred via error propagation from the standard deviations around the mean values of the density matrix elements, calculated from $N=500$ realizations of turbulence.}
\label{Fig:BellParameters}
\end{figure}

In \sref{Sec:Ent}, we showed that in the higher-dimensional cases ($d=3,4$), when no correction is employed, the entanglement transmission is almost independent of the OAM modes used to encode the information. 
Furthermore, regardless of the dimension of the encoding subspace $\mathcal{B}$, states encoded in lower OAM modes experience reduced trace losses and, in the strong scintillation regime, their entanglement is protected more efficiently by AO.
Therefore, in this section, for each dimension $d$, we only consider states encoded with low order OAM modes: $\mathcal{B} = \{-1,1\}$ for $d=2$, $\mathcal{B} = \{-1,0,1\}$ for $d=3$, and $\mathcal{B} = \{-2,-1,1,2\}$ for $d=4$.

Results for the Bell parameter $S_d$, as a function of the turbulence strength $\mathcal{W}$, are shown in \fref{Fig:BellParameters}. 
In the absence of AO corrections [\fref{Fig:BellParameters}(a)], the Bell parameter $S_d$  quickly leaves the region where the inequalities \eref{BellInequality} are violated (green shaded area). 
When the tip-tilt correction is applied [\fref{Fig:BellParameters}(b)], violations are observable up to more than twice the critical turbulence strength in the uncorrected case. 
Finally, for ideal AO correction  [\fref{Fig:BellParameters} (c)], for $d=2,3$, violations of \eref{BellInequality} prevail for all here considered turbulence strengths and $S_d$ drops below the non locality threshold only at very strong disorder for $d=4$.

One of the most interesting features in \fref{Fig:BellParameters} is the observation that the slightly stronger violations of \eref{BellInequality} achievable with higher dimensional states (as shown in \ref{BellOperator}) are completely compensated for by their faster degradation in turbulence. 
In fact, the $d=2$ results (blue dots) exhibit stronger violations for all non-vanishing turbulence strengths in the uncorrected case [\fref{Fig:BellParameters} (a)], and for most of them when corrections are applied [\fref{Fig:BellParameters} (b),(c)]. 

\section{Conclusion}
\label{Sec:conc}

In this work, we investigated the potential of adaptive optics to 
mitigate the decay of entanglement and signal of high-dimensional OAM states in atmospheric turbulence.
When only minimalistic (tip-tilt) corrections are applied, we observe an entanglement enhancement by a factor of two to four, 
depending on the scintillation strength. 
Thus, state-of-the-art AO systems, which also correct higher-order phase distortions, can achieve better results. 
An upper boundary for the performance is given by our simulations of an ideal AO, which suggest that, in the weak scintillation regime, entanglement can be either almost fully protected or, depending on the dimension of the encoding subspace, suffers a reduction of less than 20\% in the worst-case scenario. 
In contrast, in the strong scintillation regime, as expected, we observe a quick reduction of the ability of AO to prevent the entanglement decay, even in an ideal AO scenario.

Both for qutrits and ququarts, increasing the OAM of the modes in the encoding subspace does not lead to enhanced entanglement stability in the uncorrected case \cite{ZhangPRA2016}. 
Moreover, lower order OAM modes are corrected more efficiently. Thus, it seems preferable to encode information into modes with a smaller azimuthal index.
Second, because of the enhanced crosstalk between neighbouring modes with increasing scintillation strength, it may be reasonable to exclude adjacent modes from the encoding subspace \cite{ZhangPRA2016}. %

We also evaluated the capability of the transmitted states to violate generalized Bell inequalities. From our findings it is clear that higher dimensional states undergo a faster entanglement decay, and that stronger Bell inequality violations, in principle attainable for high-dimensional qudits, are unable to compensate for this effect. This observation is in agreement with the previous studies of high-dimensional Bell inequalities with noisy qudits \cite{PhysRevA.93.032130} or in presence of random perturbations acting either on the states or on the mesurement settings \cite{1367-2630-18-1-013021}. Since a violation of such an inequality represents a security proof for entanglement-based QKD, our results show that we cannot improve the security of free space quantum communication simply by increasing the dimension of the encoding subspace. Fortunately, AO corrections can restore security even in the strong scintillation regime, enabling the use of high-dimensional states and, consequently, increasing the channel capacity. In a realistic scenario, one might consider to dynamically adapt the dimensionality of the encoding subspace to the instantaneous turbulence condition, in order to always maximize the trade-off between security and key rate.

Let us finally note that here we considered Gaussian beacon beams.
In the future, it will be interesting to investigate different beacon geometries, which could provide better overlap with the distorted modes and, thereby, allow for more efficient phase corrections.

\ack
The authors acknowledge support by the state of Baden-W\"urttemberg through bwHPC. G.S., V.N.S. and A.B. acknowledge support by the Deutsche Forschungsgemeinschaft under Grant No. DFG BU 1337/17-1.  

\appendix
\section{Bell Operator}
\label{BellOperator}
\setcounter{section}{1}

In this appendix we derive a Bell operator $\hat{S}_d$ from which the Bell parameters $S_d$ can be calculated as \cite{BraunsteinPRL1992}
\begin{equation}
S_d = \Tr(\hat{S}_d \hat{\rho}).
\label{Sd}
\end{equation}
The latter are defined \cite{CollinsPRL2002} in terms of the coincidence probabilities for measurements performed by two observers Alice and Bob. Both observers have two detector settings $A_a$ and $B_b$, with $a,b = \{1,2\}$. The measurement bases associated with these settings are given by
\begin{eqnarray}
\ket{v}^A_a = \frac{1}{\sqrt{d}} \sum_{j=0}^{d-1}\exp \left[i\frac{2\pi}{d}j(v+\alpha_a) \right]\ket{j}, \label{Abasis} \\
\ket{w}^B_a = \frac{1}{\sqrt{d}} \sum_{j=0}^{d-1}\exp \left[-i\frac{2\pi}{d}j(w+\beta_b) \right]\ket{j}, \label{Bbasis}
\end{eqnarray}
where $\alpha_1 = 0$, $\alpha_2 = 1/2$, $\beta_1 = 1/4$, $\beta_2 = -1/4$ and $v,w = \{0,\cdots,d-1\}$ denote the outcomes of Alice's and Bob's measurements, respectively.
The basis states $\ket{j}$ (with $ j = 0,\cdots, d-1$) correspond to the OAM states in the encoding subspace $\mathcal{B}$. For the photon that is kept in Alice's laboratory, $j=0$ corresponds to the smallest azimuthal index, $j=1$ corresponds to the second smallest azimuthal index, and so on. For the photon that is sent to Bob through the free space link, the order is inverted. The input state \eref{MaxEntState} then reads $\sum_{j=0}^{d-1}\ket{j,j}/\sqrt{d}$, for which the bases \eref{Abasis},\eref{Bbasis} are known \cite{CollinsPRL2002} to maximize the violation of \eref{BellInequality}.

The probability that the measurement outcomes $A_a$ of Alice and $B_b$ of Bob, respectively, differ by $k$ (modulo $d$) is given by the expectation value of the operator
\begin{equation}
\hat{P}(A_a = B_b +k) = \sum_{r=0}^{d-1}\ket{r}_a^A\ket{r+k\;\mathrm{mod}\;d}_b^B ~^B_b\!\bra{r+k\;\mathrm{mod}\;d}~^A_a\!\bra{r},
\end{equation}
and the Bell operator is finally given by \cite{BraunsteinPRL1992, DadaNatPhysLett2011}
\begin{eqnarray}
\fl\hat{S}_d = \sum_{k=0}^{[d/2]-1}\left(1-\case{2k}{d-1}\right) 
&\lbrace+[\hat{P}(A_1=B_1+k) + \hat{P}(B_1=A_2+k+1) \nonumber\\
 &+ \hat{P}(A_2=B_2+k) + \hat{P}(B_2=A_1+k)] \nonumber \\
&-[\hat{P}(A_1=B_1-k-1) + \hat{P}(B_1=A_2-k)\\
&+\hat{P}(A_2=B_2-k-1) + \hat{P}(B_2=A_1-k-1)]  \rbrace. \nonumber 
\end{eqnarray}

\Table{\label{Tab:Sd} Violation of the inequalities $S_d\leq2$ for the Bell states \eref{MaxEntState} ($S_d^{Bell}$) compared with the maximum possible violation ($S_d^{max}$).}
\br
$d$ & $S^{Bell}_d$ & $S_d^{max}$\\
\mr
2 & 2.8284 & 2.8284\\
3 & 2.8729 & 2.9149\\
4 & 2.8962 & 2.9727\\
\br
\endTable

The maximum possible violation of the inequalities \eref{BellInequality} is given by the largest eigenvalue ($S_d^{max}$) of $\hat{S}_d$. 
Interestingly, the eigenstate corresponding to $S_d^{max}$, for $d>2$ is not the state \eref{MaxEntState} (see \tref{Tab:Sd}), but a pure non-maximally entangled state which is however entangled in all the $d$ modes \cite{AcinPRA2002}.

\section*{References}
\bibliography{biblio-AO-hd}
\bibliographystyle{iopart-num}
\end{document}